\documentclass{jfm}
\usepackage{graphicx}
\usepackage{comment}
\usepackage{enumitem}% This line is addc by author
\usepackage{amsmath}% This line is add by author
\usepackage{mathrsfs}% add by author
\usepackage{amsmath,bm} % This line is add by author
\usepackage{multirow} % This line is add by author
\usepackage{booktabs} % This line is add by author
\usepackage{mathtools} % This line is add by author
\usepackage{amssymb} % This line is add by author
\usepackage{natbib} % This line is add by author
\usepackage{amsfonts} % This line is add by author
\usepackage{amsgen} % This line is add by author
\usepackage{bookmark}% This line is add by author
\usepackage{hyperref}% This line is add by author
\DeclareMathOperator{\tr}{tr}
\usepackage{enumitem}
\newlist{myenumi}{description}{10}
\setlist[myenumi]{leftmargin=23pt,itemsep=2pt,topsep=2pt,parsep=2pt}
\newlist{myenumi2}{description}{10}
\setlist[myenumi2]{leftmargin=40pt,itemsep=2pt,topsep=2pt,parsep=2pt}

\makeatother
\shortauthor{Jiang, Atoufi, Zhu, Lefauve, Taylor, Dalziel \&  Linden}

\title{Geometry of stratified turbulent mixing: local alignment of the density gradient with rotation, shear and viscous dissipation}
\shorttitle{Geometry of stratified turbulent mixing}

\author{Xianyang ~Jiang\aff{1}\corresp{\email{xj254@cam.ac.uk}}, Amir Atoufi\aff{1}, Lu Zhu\aff{1}, Adrien Lefauve\aff{1}, John R. Taylor\aff{1}, Stuart B. Dalziel\aff{1}
    \and P.~F.~Linden\aff{1}}
\affiliation{\aff{1}Department of Applied Mathematics and Theoretical Physics, University of Cambridge, Centre for Mathematical Sciences, Wilberforce Road, Cambridge CB3 0WA, UK}
\begin{document}
\maketitle
% \tableofcontents

\begin{abstract}
We introduce a geometric analysis of turbulent mixing in density-stratified flows based on the alignment of the density gradient in two orthogonal bases that are locally constructed from the velocity gradient tensor. The first basis connects diapycnal mixing to rotation and shearing motions, building on the recent `rortex--shear decomposition' in stratified shear layers (Jiang et al. \textit{J. Fluid Mech.}, \textbf{947}, A30, 2022), while the second basis connects mixing to the principal axes of the viscous dissipation tensor. Applying this framework to datasets taken in the stratified inclined duct laboratory experiment reveals that density gradients in locations of high shear tend to align preferentially (i) along the direction of minimum dissipation and (ii) normal to the plane spanned by the rotex and shear vectors. The analysis of the local alignment across increasingly turbulent flows offers new insights into the intricate relationship between the density gradient and dissipation, and thus diapycnal mixing. %This, in turn, enhances our comprehension of diapycnal mixing.}
%We highlight the advantages of using such local alignment measures as proxies for mixing efficiency, as they provide more mechanistic insight and do not require the use of fluctuating variables or the sorting of the density field.
\end{abstract}

\section{Introduction}
The physics-based parameterisation of turbulent mixing in density-stratified flows is a fundamental challenge in geophysical and environmental fluid dynamics \citep{CaulfieldPRF2020,dauxois_confronting_2021}. 
This challenge requires an understanding of the small-scale mechanisms that drive transport and irreversible mixing across isopycnals \citep{Gregg2018}. 
These mechanisms rely on the interaction between the velocity and scalar fields.  
In the past, the strain rate tensor was used to analyse the geometry of dissipation in shear flows (see, e.g. \citealp{Ashurst1987}). The alignment of the principal directions of the strain rate tensor with the scalar gradient plays a crucial role in the small-scale mixing and cascade processes \citep{Garcia2005}. In the context of sheared stratified turbulence, \cite{Smyth1999} investigated how the direction of the scalar gradient aligns with the principal directions of the time varying strain rate tensor during the development of Kelvin-Helmholtz instability.%A detailed analysis was performed to examine the imperfect alignment arising due to the time varying strain field. 
Unlike previous numerical studies relying on the strain rate tensor, the present work uses three-dimensional (3-D) experimental datasets and the pseudo-dissipation tensor (to be defined in \S~2.4) to study the alignment statistics with the scalar gradient.

Using simultaneous 3-D velocity and density data in the `stratified inclined duct' (SID) laboratory experiment, \cite{Jiang2022} analysed the morphology of coherent (Eulerian) vortical structures and their evolution from pre-turbulent Holmboe waves to fully-developed sheared turbulence. They explained how turbulent hairpin vortices across the density interface engulf unmixed fluid into the stratified interface, while vortices within the shear layer generate further stirring and small-scale shear, enhancing irreversible mixing. Although significant emphasis has been placed on understanding the interaction between vortices and stratification in previous research (see also \citealp{watanabe2019}), the alignment of velocity and density gradients under different shear strengths, and the connection with viscous dissipation remain unclear.

%The  present work uses the same dataset but presents an extended analysis by exploring shear-conditioned alignment statistics, incorporating the shear-vortex basis and dissipation directions, to investigate mixing.

The objective of this paper is to tackle this by using the same datasets as \cite{Jiang2022} and quantifying the relation between kinematic rotational and shearing flow structures and the dynamics and energetics of the flow, an open question \citep{Riley2022}  essential to improve parameterisations. 
Specifically, we will address the two following questions. 
%Firstly, how is viscous dissipation connected to shear and rotation (the two building blocks of vorticity) having dynamically distinct roles on mixing? 
First, how is viscous dissipation connected to the homogenisation of the density field which indicates  mixing? %whose homogenisation signals mixing? 
Second, how do shear and rotation, the building blocks of vorticity, contribute distinctively to the mixing of the density field? 
To answer these questions, we investigate the local geometric alignment of the density gradient with characteristic directions associated with the local viscous dissipation tensor and with the rotation and shear vectors.
%the which plays a role in the amplification of the strain rate tensor \citep{Buaria2022b}. \AL{we dont really explain why the sst is important or even study it in this paper, so why cite this paper here?}

%Previous studies have shown that vorticity tends to align with the intermediate eigenvector of the strain rate tensor in isotropic turbulence \citep{Ashurst1987}. \AL{This is between vorticity and strain rate so not related to density field, irrelevant to this paper}

% Irreversible mixing can be quantified globally using the available and background potential energy decomposition of \cite{Winter1995}, but doing so requires access to and sorting of the entire three-dimensional density field into a statically stable state, which poses challenges for laboratory or field data. Alternatively, mixing parameters can be defined locally from more easily accessed quantities such as the ratio of the buoyancy flux or the rate of scalar variance dissipation to the rate of turbulent kinetic energy dissipation \citep{CaulfieldPRF2020}. However, these energy fluxes are based on the fluctuating (or `turbulent') velocity and density fields, which poses theoretical and practical challenges around the averaging procedure, especially in spatio-temporally intermittent flows. To overcome these challenges, we propose a third and new approach based on the local geometric alignment of the density gradient with suitably defined directions of viscous dissipation, and we use total rather than fluctuating quantities.

In \S\,\ref{sec:method} we introduce our experimental datasets and new geometric framework consisting of two local orthogonal bases constructed from the velocity gradient tensor. We quantify the alignment of the density gradient in the first basis (spanned by viscous dissipation eigenvectors) in \S\,\ref{sec:egradrho}, and in the second basis (spanned by shear and rotation vectors) in  \S\, \ref{sec:align_structure_gradrho}. We then study in \S\,\ref{sec:Decomp_densitygradient_dissip} how this alignment varies under increasing turbulent intensity, and in \S\,\ref{sec:ImplicationMix} examine the link between alignment and a standard mixing coefficient in the most turbulent flow, emphasising the physical insights gained. Finally, we conclude in \S\,\ref{sec:Con}.

\section{Methodology}
\label{sec:method}

\subsection{Experiment and data processing}

\begin{figure}
\centering
\includegraphics[width=0.95\textwidth]
{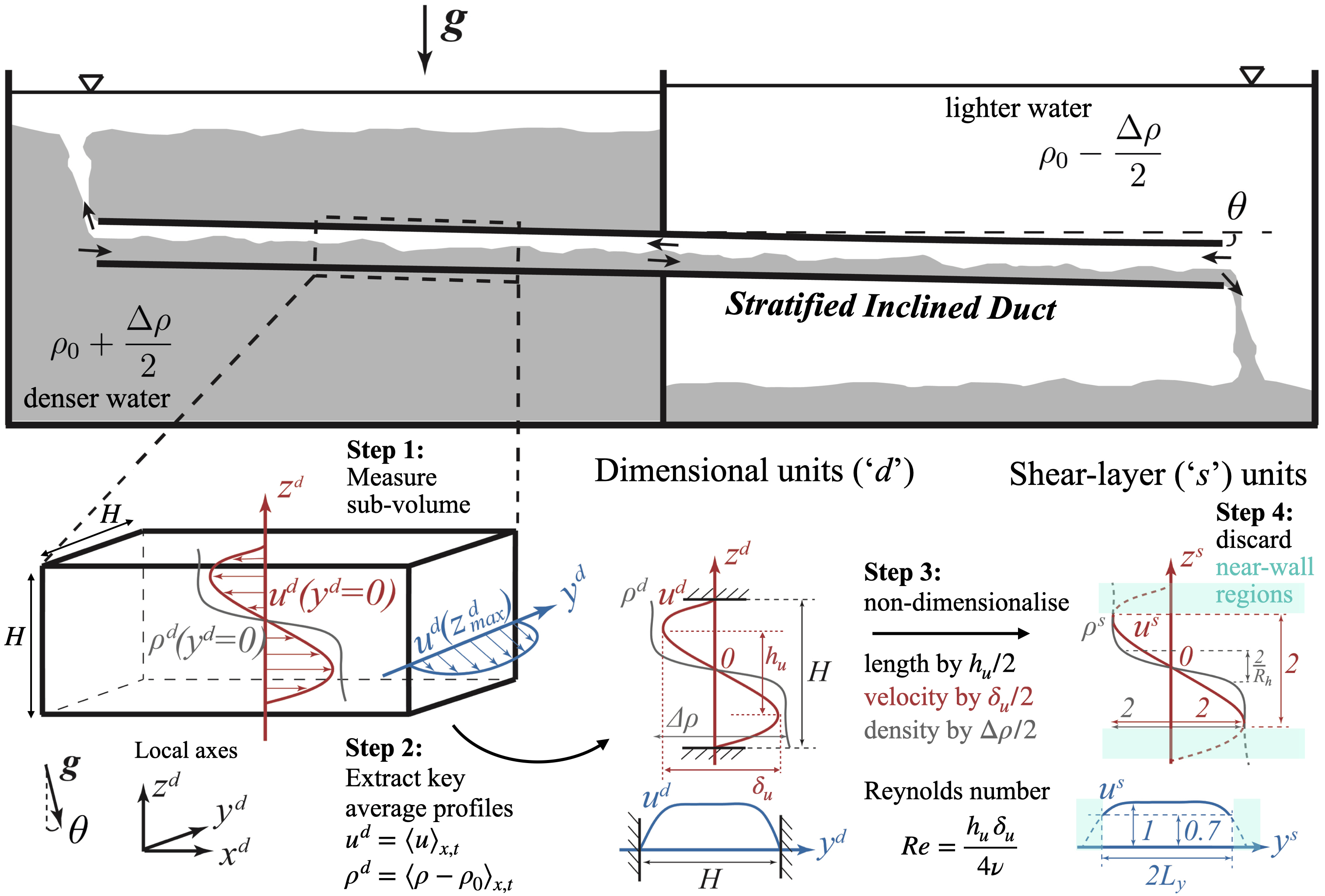}% 
\caption{ Setup with the four-step data acquisition and processing pipeline, transforming volumetric velocity and density measurements into non-dimensional `shear-layer' datasets with a peak-to-peak velocity jump, density jump, and shear layer heights of 2.}
\label{fig:setup}
\end{figure}

We collected data in the stratified inclined duct (SID) experiment (sketched in figure~\ref{fig:setup})  in which a salt-stratified, shear-driven flow is sustained in a square duct of length $1350$ mm and cross-section $H=45$ mm (aspect ratio $30$) connecting two reservoirs at different densities $\rho_0\pm\Delta\rho/2$ , giving an Atwood number $At \equiv \Delta\rho/(2\rho_0)$. The duct can be tilted at a small angle $\theta$ with respect to the horizontal to add gravitational forcing and increase turbulent levels. This experiment has been described in more detail in \cite{meyer2014} and \cite{lefauve2020}. The data acquisition and processing pipeline, originally introduced in \cite{Lefauve20221} (\S~3.3), comprises four steps illustrated in figure~\ref{fig:setup}.

In step 1,  we simultaneously measure the time-resolved, 3-D density field $\rho^d(\mathbf{x}^d,t^d)$ and three-component velocity field $\bm{u}^d=(u^d,v^d,w^d)(\mathbf{x}^d,t^d)$ using scanning laser-induced fluorescence and stereo particle image velocity, \citep{partridge2019}. The superscript $d$ highlights that the data are still in the original dimensional units. The $x^d$-axis is aligned locally with the duct, making the $z^d$-axis tilted at an angle $\theta$ with respect to the true vertical (defined by the direction of gravity $\mathbf{g}$), while the $y^d$-axis is the spanwise direction. 

In step 2, the velocity $u$ and density $\rho-\rho_0$ are then averaged in $x^d$ and in $t^d$, yielding three key profiles: the vertical profile of the streamwise velocity, $u^d(z^d)$ (in red), sampled at $y^d=0$, the vertical profile of density, $\rho^d(z^d)$ (in grey), also sampled at $y^d=0$, and the spanwise profile of the streamwise velocity, $u^d(y^d)$ (in blue), sampled at the height corresponding to maximum flow speed, $z^d_{\text{max}}$.

In step 3,  all data are then non-dimensionalised. For length we use half the shear layer height $h_u/2$ (measured from the locations of minimum to maximum $u^d(z^d)$), for velocity we use half the peak-to-peak velocity jump of $u^d(z^d)$, and for density we use half the fixed maximum density jump $\Delta \rho/2$. This restricts all non-dimensional shear-layer variables $z^s,u^s,\rho^s$ between $-1$ and $1$, and is consistent with the following Reynolds and bulk Richardson numbers: 
\begin{equation} \label{eq:nondim}
   Re \equiv \Big( \frac{h_u}{2} \, \frac{\delta_u}{2}\Big)\Big/\nu, \quad Ri_b \equiv \frac{g}{\rho_0} \frac{\Delta\rho}{2}\frac{h_u}{2}\Big/\Big(\frac{\delta_u}{2}\Big)^2. 
\end{equation}
Note that salt stratification yields a large Prandtl number $Pr\equiv \nu/\kappa \approx 700$, where $\nu$ is the kinematic viscosity of water and $\kappa$ the diffusivity of salt. The thickness of the density interface, defined as the spacing between the values $\rho^d=\pm \tanh(1) = \pm 0.76$, becomes $2/R_h$, where $R_h$ is the thickness ratio between the shear layer and the density interface. This step allows datasets obtained in flows at different values of $\Delta \rho$ (or Atwood number $At$) and tilt angle $\theta$ to be compared side-by-side with accurate non-dimensional parameters defined based on the actual measured velocity profiles.

In step 4, we define the shear layer as the region $|z^s|\le 1$ and $|y^s|\le L_y$ (where $2L_y$ is the central portion where $|u^s|>0.7$) and discard the region outside this. %regions outside the shear layer are cropped (discarded), defined as $|z^s|>1$ and $|y^s|>L_y$ where $2L_y$ is the central portion where $|u^s|>0.7$. 
This step avoids including  wall-bounded, unstratified flow (inherent to this experiment) in our statistical analysis of the interfacial, shear-driven stratified flow data of interest to this paper. In the remainder of the paper, we drop the $s$ superscripts and use shear layer variables throughout.

\subsection{Dataset}
 
The dataset, described in \cite{Lefauve20221} and freely available in \cite{lefauve2022dataset}, includes 15 experiments with increasing levels of turbulent intensity, controlled by the product of the tilt angle $\theta$ and Reynolds number $\Rey$. The data comprise four flows in the Holmboe regime (H1--H4), showing small-scale, travelling waves whose scouring motion maintain a relative sharp interface $R_h\in [7.2, \, 11.3]$; eight flows in the intermittently turbulent regime (I1--I8) with bursts of increasingly longer-lived turbulence whose overturning mixing results in a thicker interface ($R_h \in [2.2,\, 5.8]$); and three flows in the turbulent regime (T1--T3), where the flow never relaminarises and the density interface is thickest ($R_h\in [1.8,\, 2.1]$). 

Our analysis focuses on two typical cases: H1, a Holmboe wave flow at $(\theta,\Rey,Ri_b,R_h)=(1^\circ, \,381,\, 0.567,\,8.9)$; and T3, a vigorously turbulent flow at $(\theta,\Rey,Ri_b,R_h)=(5^\circ,\, 1145, \, 0.147, \, 1.9)$. For detailed parameters of these and the remaining datasets used in this paper, see \cite{Lefauve20221}, Table 1.

%, while $\rho=0$ represents neutral buoyancy. 
%The mean flows in both cases can be approximated by a stratified shear layer described by $u=-\tanh ,z$ and $\rho=-\tanh , R_h z$, with some variation along the $y$ direction. 
%The resolution of the data is given by $\Delta x$, $\Delta y$, and $\Delta z$.

% \begin{table}
%   \begin{center}
% \def~{\hphantom{0}}
% \setlength{\tabcolsep}{6pt}
%   \begin{tabular}{ccccccccc}
% Case &  \multicolumn{5}{c}{Parameters}  & \multicolumn{3}{c}{Resolution}   \\[3pt]
%  & $Pr$ & $\theta$ ($^\circ)$ & $Re$   &   $Ri_b$ & $R_h$& $\Delta x$, $\Delta z$ & $\Delta y$ & $\Delta t$ \\[3pt]
% H1   & 700 & 1 &   381   &    0.567 &   8.9 &  0.047  &  0.095   &  2.05   \\[5pt]
% T3     & 700 &  5   &   1145            &   0.147   &   1.9    &    0.036  &       0.073     &   3.70   \\
%   \end{tabular}
%   \caption{Properties of the Holmboe (H1) and turbulent (T3) data sets primarily used in the paper, adapted from \cite{Lefauve20221}'s tables 1 and 3.}
% \label{tab:dataset}
%   \end{center}
% \end{table}

%%%%%%%%%%%%%%%%%%%%%%%%%%%%%%%

\subsection{Vorticity decomposition and structural coordinates} 
\label{sec:RSdecomp}

We build on the kinematic analysis of \cite{Jiang2022} using the local and instantaneous rortex-shear decomposition of vorticity $\bnabla \times \bm{u} \equiv \bm\omega = \bm{R}+\bm{S}$, appropriate when $\bnabla\bm{u}$ has only a single real eigenvalue. The `rortex vector', capturing rigid-body rotation, is defined as \citep{Gao2019}
\begin{equation}
    {\bm{R}\equiv \left(1-\sqrt{1-\frac{4\lambda_{ci}^2}{(\bm\omega\cdot\hat{\bm r})^2}}\right)(\bm\omega\cdot\hat{\bm r})\hat{\bm r}} = R \, \hat{\bm r},
    \label{RotexEqn}
\end{equation}
where $\hat{\bm r}$ and $\lambda_{ci}$ are, respectively, the local unit \emph{real} eigenvector and \emph{imaginary} part of the complex-conjugate pair of eigenvalues of the velocity gradient tensor $\bnabla\bm{u}$. Points where such a complex-conjugate pair does not exist have no local rigid-body rotation and are thus excluded from our analysis. The remaining vorticity $\bm{S}\equiv \bm \omega - \bm R$ is the called `shear vector' (with magnitude $S$) as it is dominated by shearing motions (especially in SID where turbulence is fed by a sustained background shear). Since the shear $\bm{S}$ is not necessarily perpendicular to $\bm R$, it can be further decomposed into `rotational shear' $\bm{S}_r=(\bm{S} \cdot \hat{\bm r})\hat{\bm r}$ (aligned with $\bm R$) and a residual `non-rotational shear' $\bm{S}_{n}$, as shown in figure \ref{fig:RSdecomp}(\textit a). We define the total rotation as $\bm{R}_t \equiv \bm{R} +  \bm{S}_r$ and construct the right-handed orthonormal basis $(\hat{\bm{n}},\hat{\bm r},\hat{\bm f})$ based on the unit non-rotational shear and rotation vectors
\begin{eqnarray}
\hat{\bm n}= \frac{\bm{S}_{n}}{|\bm{S}_{n}|}, \quad \hat{\bm r} = \frac{\bm{R}_{t}}{|\bm{R}_{t}|}, \quad  \hat{\bm{f}} \equiv \hat{\bm{n}} \times \hat{\bm{r}}.
\end{eqnarray}
%
% Figure \ref{fig:RSdecomp}(\textit b) illustrates the typical flow motions induced corresponding to $\bm r$ (in red), $\hat{\bm n}$ (in green) and their effect on three hypothetic isopycnal planes $\rho=\rho_1$, $\rho_2$ and $\rho_3$ (in grey), normal to $\bm r$, $\hat{\bm n}$, and  $\hat{\bm f}$, respectively.

%%%%%%%%%%%%%%%%%%%%%%%%%

\subsection{\label{sec:DensityDecomp}Directions of dissipation } 

%\AL{Make link with Cauchy-Green tensor in solid mechanics and polar decomposition? Direction of most stretch, Quadratic form interp. Lyapunov exponent, perhaps direction of strain (meaning of the e-vecs of $V$ in the polar decomposition. }
%\XJ{May relate the work rate done and dissipation rate produced by various stress constituents}
%\AL{I made some changes, and I think this might be enough to 'justify' the interpretation as max, intermediate and min dissipation}

\begin{figure}
\centering
\includegraphics[width=1.0\textwidth]
{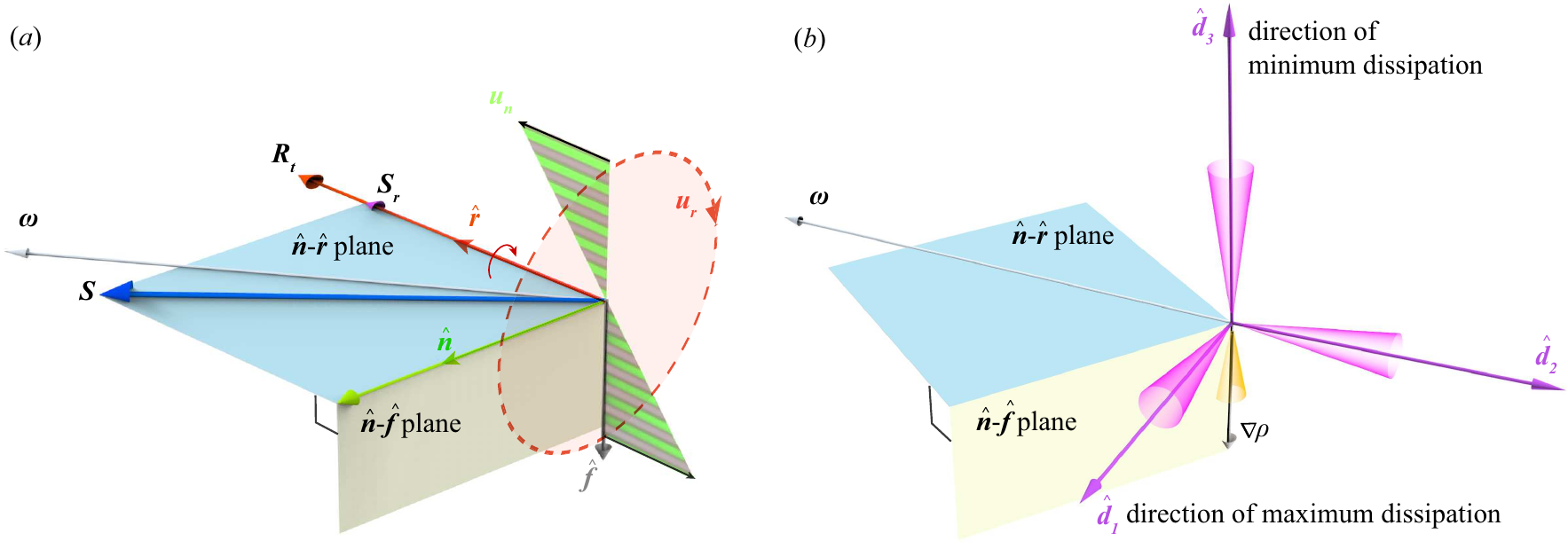}% 
\caption{Sketch of the orthogonal bases  (\textit a) $(\hat{\bm n}, \hat{\bm r}, \hat{\bm f})$ based on  rigid-body rotation $\hat{\bm r}$ in the velocity gradient tensor and non-rotational vorticity $\hat{\bm n}$; (\textit b) $(\hat{\bm{d}_1},\hat{\bm{d}_2},\hat{\bm{d}_3})$ based on the eigendirections of the pseudo-dissipation tensor. In (\textit a), red dashed arrow labelled $\bm u_r$ represents velocity induced by rigid-body rotation and the green band labelled $\bm u_n$ represents the likely velocity distribution induced by the non-rotational shear. % the velocity corresponding to $\hat{\bm r}$ (respectively $\hat{\bm n}$) is represented by a red dashed arrow (respectively a green band).
In (\textit b), the shaded cones represent the typical range of alignments found in our results.}
\label{fig:RSdecomp} 
\end{figure}

To gain dynamical information, we define the local and instantaneous pseudo-dissipation tensor (analogous to the left Cauchy--Green tensor in solid mechanics): %\AL{Consider calling the pseudo diss tensor $\mathsfbi D$ and dissipation directions $\bm d_i$ as it emphasises the meaning}   %\AL{We should probably have this after RS as it is 'new' and we should say why we introduce this rather than just sticking with R-S} 
\begin{equation}\label{eq:pseudo-diss-tensor}
    \mathsfbi D \equiv  \frac{2}{\Rey}  \bnabla\bm{u} \cdot \bnabla\bm{u}^T \quad\text{or} \quad D_{ij} \equiv \frac{2}{\Rey}  \frac{\partial u_i}{\partial x_k}\frac{\partial u_j}{\partial x_k}.
\end{equation}
There is a close relation between the rate of kinetic energy dissipation typically used in the stratified turbulence literature $\epsilon\equiv (2/\Rey) \mathsfbi E : \mathsfbi E$ (where $\mathsfbi E \equiv \left(\bnabla\bm{u} + \bnabla\bm{u}^T \right)/2$ is the strain rate tensor)  and the pseudo-dissipation $\tilde{\epsilon}\equiv (1/\Rey) \bnabla \bm{u} : \bnabla \bm{u}$  \citep[\S~5.3]{Pope2000}, namely 
\begin{equation}
\tilde{\epsilon} = \frac{1}{2} \tr({\mathsfbi D}) = \frac{1}{\Rey} \dfrac{\partial u_i}{\partial x_j} \dfrac{\partial u_i}{\partial x_j} = \epsilon  - \frac{1}{\Rey} \dfrac{\partial u_i}{\partial x_j} \dfrac{\partial u_j}{\partial x_i} \approx \epsilon.  
\end{equation}
%
% \begin{eqnarray}
% \epsilon \equiv \frac{2}{\Rey} \ E_{ij} E_{ij} = \frac{1}{\Rey} \dfrac{\partial u_i}{\partial x_j} \dfrac{\partial u_i}{\partial x_j} + \frac{1}{\Rey}  \dfrac{\partial u_i}{\partial x_j} \dfrac{\partial u_j}{\partial x_i}= \tilde{\epsilon}+ \frac{1}{\Rey} \dfrac{\partial u_i}{\partial x_j} \dfrac{\partial u_j}{\partial x_i}\approx \tilde{\epsilon},
% \end{eqnarray}
% %
Our motivation for using $\mathsfbi{D}$ (rather than $\mathsfbi{E}$) to study dissipation in this paper is that $\mathsfbi{D}$ is closely related to the eigenbasis of  $\bnabla\bm{u}$, upon which structural identification schemes are often built, like the above rortex-shear decomposition. The advantage of  $\mathsfbi{D}$ is that (unlike $\bnabla\bm{u}$) it is symmetric positive semi-definite and therefore has three real positive eigenvalues $\lambda_1 \ge \lambda_2 \ge \lambda_3 \ge 0$, such that $\tilde{\epsilon}=\frac{1}{2}(\lambda_1+ \lambda_2+\lambda_3)$,  %\AL{To say they are all positive you need to prove e is semi-positive definite}
% \begin{equation} \label{eq:pseudo-diss-as-sum-of-evals}
%     \tilde{\epsilon}=\frac{1}{2}(\lambda_1+ \lambda_2+\lambda_3),
% \end{equation}
%Since all three second-order and first-order principal minors of $\mathsfbi{D}$ are nonnegative, therefore $\mathsfbi{D}$ is positive semidefinite due to the generalized Sylvester criterion. This also implies that the eigenvalues of $\mathsfbi{D}$ are nonnegative. 
and an associated basis of real, orthogonal unit eigenvectors $(\hat{\bm d_1},\hat{\bm d_2}, \hat{\bm d_3})$, sketched in pink in figure \ref{fig:RSdecomp}(\textit b). 

The interpretation of this basis is provided by the unique polar decomposition $\bnabla\bm{u}=\mathsfbi{V} \cdot \mathsfbi{R}$, 
where  $\mathsfbi{R}$ is the rotation tensor (a proper-orthogonal tensor) and $\mathsfbi{V}$ is the left stretch tensor (a real, symmetric, positive semi-definite tensor; \citealp[\S~9.2]{spencer_continuum_1980}). Since $\mathsfbi{R}$ is orthogonal then $\mathsfbi{D}=\frac{2}{\Rey}\mathsfbi{V}^2$ and $\mathsfbi{V}$ has the same eigenvectors as $\mathsfbi{D}$ with associated eigenvalues $\sqrt{\Rey\lambda_i/2}$. The linear transformation $\bm{\delta u}=\bnabla\bm{u} \cdot \bm{\delta x}$ (the first order change of velocity a small $\bm{\delta x}$ away) is decomposed into a rotation followed by a local stretch (if $\sqrt{\Rey\lambda_i/2}>1$) or compression (if $\sqrt{\Rey\lambda_i/2}<1$) along the basis vectors $\hat{\bm d_i}$. As the dissipation equals (half) the sum of these principal stretches squared, % \eqref{eq:pseudo-diss-as-sum-of-evals}, 
we interpret $\hat{\bm d_1},\hat{\bm d_2}, \hat{\bm d_3}$ as representing the directions of maximum, intermediate and minimum dissipation, respectively. 

According to \cite{Wu2016}, the dissipation is related to the stress resulting from the surface deformation rate, which constitutes a viscous resistance to changes in the direction and area of surface elements. Therefore, the maximum (minimum) dissipation direction can be seen as the axis along which fluid elements can efficiently (inefficiently) dissipate kinetic energy through cumulative effects of stretching (or compression) and rotation.

\subsection{Density-gradient ratios}\label{SDG}
We define the `dissipation angles' as $\zeta_i \equiv \angle(\bnabla\rho,\hat{\bm d_i})$, with $i=1, 2, 3$, and the `structure angles' as $\zeta^\phi \equiv \angle(\bnabla\rho,\hat{\bm \phi})$, with $\hat{\bm \phi}=\hat{\bm n},\hat{\bm r}, \hat{\bm f}$.
The density gradient can then be decomposed either along the directions of dissipation defined by the unit vectors $(\hat{\bm{d}_1},\hat{\bm{d}_2},\hat{\bm{d}_3})$, 
\begin{eqnarray}
\bnabla\rho = \sum_{i=1,2,3} \bnabla{\rho}_i, \quad \text{where} \quad \bnabla{\rho}_i = |\bnabla{\rho}|\cos(\zeta_i)\hat{\bm{d}_i}, 
%\quad \bnabla{\rho}_i = (\bnabla{\rho} \cdot \bm d_i)\bm{d}_i. 
\label{eqn:densityprojection1i}
\end{eqnarray}
or  along the structural direction  as 
\begin{eqnarray}
\bnabla\rho = \sum_{\phi=n,r,f} \bnabla{\rho}^\phi, \quad \text{where} 
\quad \bnabla{\rho}^\phi = |\bnabla{\rho}|\cos(\zeta^\phi)\hat{\bm{\phi}}.
\label{eqn:densityprojectionphi}
\end{eqnarray}
In (\ref{eqn:densityprojection1i}), each component $\bnabla{\rho}_i$ can be further decomposed along $(\hat{\bm n},\hat{\bm r},\hat{\bm f})$ %, creating nine vectors
\begin{eqnarray}
\bm \nabla \rho_{i}= \sum_{\phi=n,r,f} \bm \nabla \rho_{i}^\phi,  \quad \text{where} \quad  \bm \nabla \rho_{i}^\phi = (\bm\nabla \rho_{i} \cdot \hat{\bm \phi})\hat{\bm \phi}. %\ \ \text{for} \ \ \bm \phi = \bm n,\bm r,\bm f.
\label{eqn:densityprojection1j}
\end{eqnarray}

The twelve components in (\ref{eqn:densityprojection1i}) and (\ref{eqn:densityprojection1j}) encode the alignment of the density gradient with six physically meaningful directions corresponding to dissipation and rotation/shear. We propose the following $3+3+9=15$ squared-density-gradient ratios (SDGRs):
\begin{eqnarray}
\mathscr{M}_{i} \equiv \frac{|\bm \nabla \rho_i|^2}{|\bm \nabla \rho|^2} = \cos^2\zeta_i, \quad  \mathscr{M}^\phi \equiv  \frac{|\bm \nabla \rho^\phi|^2}{|\bm \nabla \rho|^2} = \cos^2\zeta^\phi, \quad 
\mathscr{M}_{i}^\phi \equiv  \frac{|\bm \nabla \rho^\phi_{i}|^2}{|\bm \nabla \rho_{i}|^2}.
\label{eqn:densityprojectionratioij}
\end{eqnarray}
%with the `dissipation angles' $\zeta_i \equiv \angle(\bnabla\rho,\bm d_i)$ and the `structure angles' $\zeta^\phi \equiv \angle(\bnabla\rho,\bm \phi)$. 
%\XJ{Here, and in the following, the summation convention over repeated indices is not applied; where summation is required, we indicate this explicitly using $\sum_j$ to indicate summation over the index $j$.} 
Here, repeated indices do not imply summation; where summation is required, we use explicitly $\sum_j$.
%The matrix $\mathsfbi{M}\equiv(\mathscr{M}_{i}^\phi)_{i=1,2,3}^{\phi=n,r,f}$ containing 
The last nine SDGRs encode the projection of $(\hat{\bm d_1},\hat{\bm d_2},\hat{\bm d_3})$ in $(\hat{\bm n},\hat{\bm r},\hat{\bm f})$ and is related to the rotation matrix $\mathsfbi{T}_{i\phi}$ between these bases. 
The use of orthonormal bases implies that each group of SDGR sum to one, namely $\sum_{i} \mathscr{M}_i = \sum_{\phi} \mathscr{M}^\phi = \sum_{\phi} \mathscr{M}^\phi_1= \sum_{\phi} \mathscr{M}^\phi_2 = \sum_{\phi} \mathscr{M}^\phi_3 = 1$.

% \subsection{Link with mixing efficiency}

% We note the following relation between $\mathscr{M}_i$ and the local mixing efficiency coefficient $\Gamma_\chi\equiv \chi/\tilde{\epsilon}$, defined here using the total (rather than fluctuating) dissipation of kinetic energy $\tilde{\epsilon}$ and scalar variance $\chi \equiv 2/(Re Pr) |\bnabla\rho|^2$ where $Pr$ is the Prandtl number.  By analogy with $\mathsfbi D$, we can define the scalar variance dissipation tensor as the dyadic product of $\bnabla{\rho}$ with itself:
% \begin{eqnarray}
%     \mathsfbi{X} \equiv \frac{2}{Re \,Pr} \bnabla{\rho} \otimes \bnabla{\rho}^T \ \ \text{or} \ \  \text{X}_{ij} \equiv \frac{2}{Re\, Pr} \frac{\partial \rho}{\partial x_i}\frac{\partial \rho }{\partial x_j}, \ \ \text{hence} \ \ \chi = \frac{1}{2}\tr{(\mathsfbi{X})} \qquad
% \end{eqnarray}
% and therefore $\Gamma_\chi=\tr{(\mathsfbi{X})}/\tr{(\mathsfbi{D})}$. The rank-one dyad $\mathsfbi{X}$ has a single non-zero eigenvalue $\abs{\bnabla \rho}^2$ and eigenvector $\bnabla{\rho}$. \AL{I don't get the following. If we can't make this statement, then what was the purpose of defining $\mathsfbi{X}$?} We thus interpret mixing as being most efficient when the dominant eigenvector of  $\mathsfbi{D}$ (i.e. $\bm{d}_1$) and $\mathsfbi{X}$ (i.e. $\bnabla \rho$) are aligned, suggesting that $\Gamma_\chi$ is inversely proportional to  $\mathscr{M}_1$ (with a factor $Pr^{-1}$).  

%%%%%%%%%%%%%%%%%%%%%%%%%
%%%%%%%%%%%%%%%%%%%%%%%%%%

\section{Results}\label{sec:results}

\subsection{Alignment of density gradient with directions of dissipation}\label{sec:egradrho}

%To get more insight into the properties of the dissipation tensor in the stratified shear layer, we can 
We start by examining the local alignment between $\bnabla\rho$ and $\hat{\bm{d}_1},\hat{\bm{d}_2},\hat{\bm{d}_3}$ quantified by the three angles $\zeta_i$. Since $\cos^2{\zeta_2} =1-\cos^2{\zeta_1} - \cos^2{\zeta_3}$, we focus our analysis on the two independent angles $\zeta_1$ and $\zeta_3$.

\begin{figure}
\includegraphics[width=0.99\textwidth]{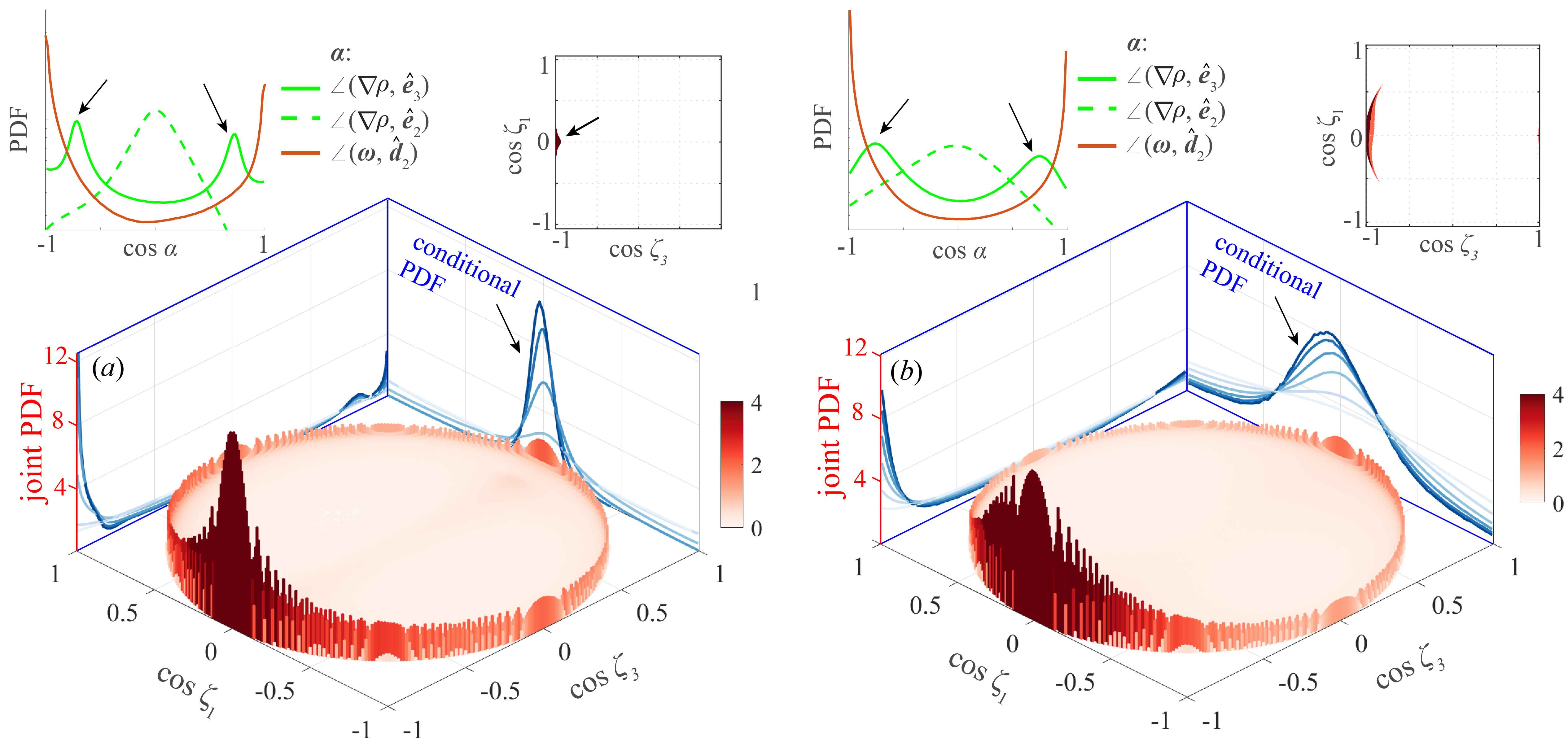}%
\caption{Joint PDFs of the alignment between $\bm \nabla\rho$ and the direction of maximum and minimum dissipation $\hat{\bm d_1}$, $\hat{\bm d_3}$ (angles $\zeta_1,\zeta_3$  respectively) in H1 (\textit a) and T3 (\textit b). Conditional PDFs  are shown in blue, with darker lines indicating higher shear thresholds $k=S/S_\textrm{rms}\in (0,0.5],  \ldots, (2.5,3]$. The vertical scale of blue lines has been amplified by  3 (a) and 1.5  (b) for better visualisation. The top-right insert shows the joint PDF conditioned at $k\in (2, 2.5]$. The top-left insets compares the alignment of $\bm \nabla\rho$ with the intermediate and minimum strain directions ($\hat{\bm e_2}$, $\hat{\bm e_3}$) to the alignment of $\hat{\bm d_2}$ with $\bm \omega$.} 
\label{fig:jpdfH1T3}
\end{figure}

Figure \ref{fig:jpdfH1T3} compares the joint probability density function (PDF) of $\cos\zeta_1$ and $\cos\zeta_3$ in the Holmboe wave flow (dataset H1, panel (\textit{a})) and turbulent flow (dataset T3, panel (\textit{b})). The blue lines represent conditional PDFs for the two angles separately at increasingly large values of $k=S/S_\textrm{rms}$, where $S_\textrm{rms}$ is the root-mean-square shear averaged in space and time, starting with the interval $(0,0.5]$ (lighter shade) and ending with $(2.5,3]$ (darkest shade). Note that the high-shear region often corresponds to a higher dissipation (dominated by $\lambda_1$). 
We find that, in both flows, $\bnabla \rho$ is most often nearly anti-parallel to $\hat{\bm d_3}$ ($\cos \zeta_3 \approx -1$) and nearly perpendicular to $\hat{\bm d_1}$ ($\cos \zeta_1 \approx 0$) and, therefore, nearly perpendicular to $\hat{\bm d_2}$.
This trend is clear even in the turbulent flow, despite more spread around the mean values than in the Holmboe flow.

The conditional PDFs and the top-right insets, restricting the joint PDFs to high-shear regions $k\in (2,2.5]$, show that this trend increases in regions of high shear.  Figure \ref{fig:RSdecomp} illustrates (with shaded cones) this preferential alignment of the density gradient $\bnabla\rho$ along the (negative) direction of least dissipation  $\hat{\bm d_3}$ and thus perpendicular to $\hat{\bm d_1},\hat{\bm d_2}$.  

The top-left insets of figure \ref{fig:jpdfH1T3} shows the PDF of the alignment of $\bnabla\rho$ with the intermediate and minimum eigendirections of the strain rate tensor ($\hat{\bm{e}_2}$ and $\hat{\bm{e}_3}$). % These directions correspond to intermediate and minimum eigenvalues of the strain rate tensor $\mathsfbi{E}$. 
It shows a preferential $\bnabla\rho \perp \hat{\bm{e}_2}$ (similarly to $\hat{\bm{d}_2}$) but a deviation from alignment to $\hat{\bm{e}_3}$ by $\approx 40^\circ$ (see black arrows). % show that $|\cos{\angle(\bnabla\rho,\hat{\bm e_3})|\approx0.75}$) 
This observation aligns with prior findings in sheared, stratified flow, which have shown imperfect alignment between $\bnabla \rho$ and the compressive strain direction \citep{Smyth1999, Ashurst1987}.  % This behaviour contrasts with isotropic flows, where scalar gradients typically align with $\hat{\bm{e}_3}$.
% The imperfect alignment indicates that the strain in the continuously shear-driven SID flow ($Pr=700$) intensify the density gradient. 
% may be attributed to the presence of side-wall confinement in the SID setup and the use of a high Prandtl number in this study. 
This finding  implies that both maximum stretching and compressive strain influence the orientation of $\bnabla\rho$ in the continuously shear-driven, high-$Pr$ SID flow, promoting alignment with the direction of minimal dissipation.

% \revise{The findings that the $\bnabla \rho$ predominantly aligns with the third direction of the dissipation tensor echo those of \cite{Smyth1999} who used the strain rate tensor, especially in the region of high shear. %This resemblance is particularly pronounced in cases with high Prandtl numbers and in the region of high shear.
% This implies that the strain field at the high-shear region influences the orientation of $\bnabla \rho$, encouraging alignment with the minimal dissipation direction, which coincides with the most compressive strain direction.} 
% \revise{The broadening and sharpening of the density interface along this direction is an essential aspect of small-scale mixing.} 

Isopycnals oriented in the directions of maximum and intermediate dissipation (i.e. $\bnabla\rho\parallel\hat{\bm d_3}$) are particularly susceptible to stretching and diffusion. The reduced alignment between $\bnabla\rho$ and $\hat{\bm d_3}$ in T3,  in contrast with H1, %as shown in figure \ref{fig:jpdfH1T3}(\textit b), 
suggests increased turbulence-induced energy dissipation, resulting in a more effectively mixed shear layer. Therefore, the deflection of $\bnabla\rho$ from the $\hat{\bm d_3}$ direction to the $\hat{\bm d_1}-\hat{\bm d_2}$ plane is an outcome of this mixing process.

We also find that the vorticity $\bm{\omega}$ has a strong and robust preferential alignment along $\hat{\bm d_2}$ in both flows (see top-left inserts of figure \ref{fig:jpdfH1T3}). This echoes earlier studies that observed vorticity aligning with the intermediate principal direction $\hat{\bm e_2}$ of  the strain rate tensor in both shear and isotropic flows \citep{Ashurst1987}.  %with this alignment strengthening as vorticity intensity increases in isotropic turbulence \citep{Buaria2020}.
It implies that the intermediate directions of the strain and dissipation tensors may share comparable functions related to the net production of enstrophy. 
%However, further work is required to explore the enstrophy-dissipation correlation in stratified shear flows.}

%%%%%%%%%%%%%%%%

\subsection{Alignment of density gradient with rotation and shear}
\label{sec:align_structure_gradrho}

\begin{figure}
\includegraphics[width=0.97\textwidth]{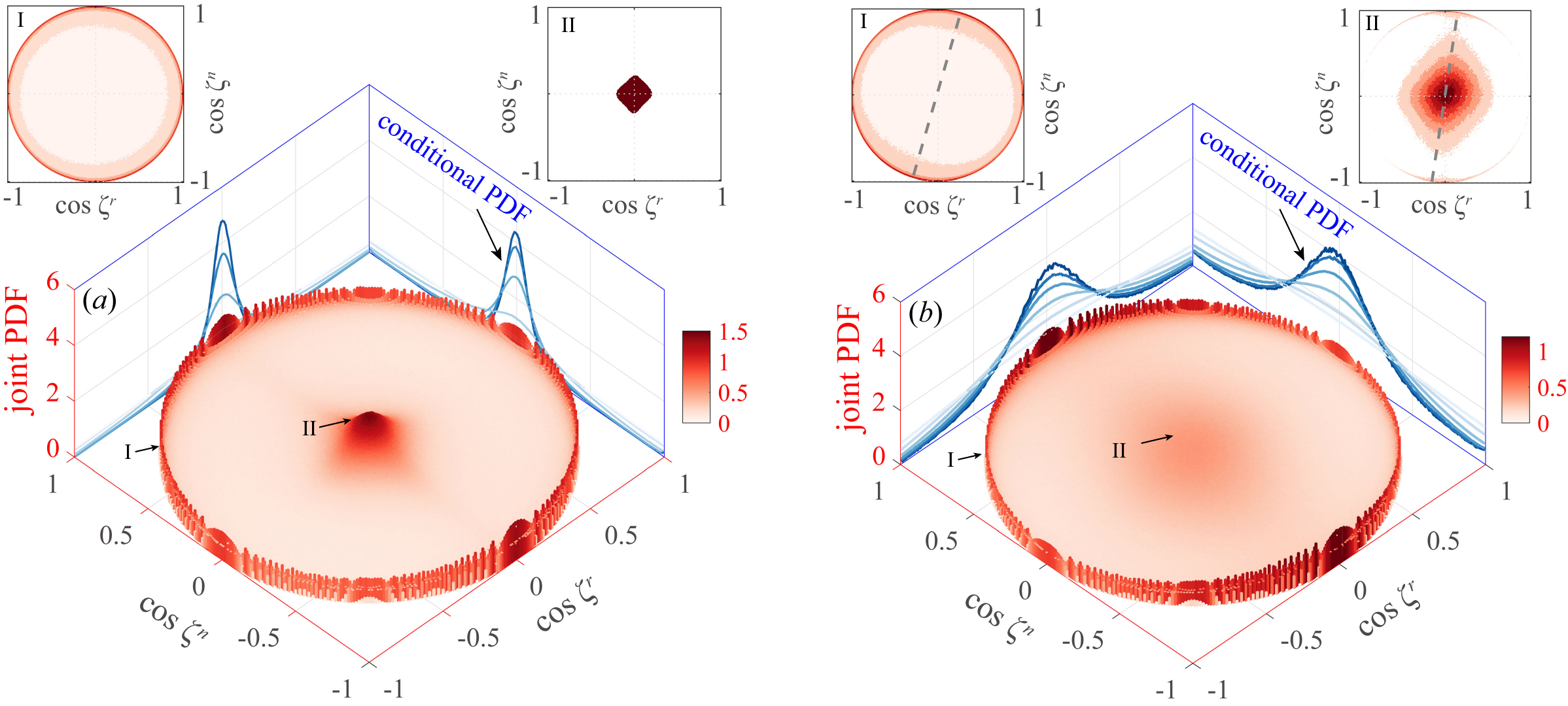}%
\caption{Joint PDFs of the alignment between $\bm \nabla\rho$ and non-rotational shear $\hat{\bm n}$ (angle $\zeta^n$) and rotation $\hat{\bm r}$ (angle $\zeta^r$)  for H1 (a) and T3 (b),  similar to figure~\protect{\ref{fig:jpdfH1T3}}. The top-left and top-right insets show the joint PDF conditioned at $k\in(0,0.5]$ and $k\in (2, 2.5]$, respectively, with the dashed line in (b) indicating the asymmetry of the distribution. }
\label{fig:Gradrho_RtSnrF_H1T3}
\end{figure}

We now examine the local alignment between $\bnabla\rho$ and the rotation/shear basis using the two independent angles $\zeta^r,\zeta^n$ in figure \ref{fig:Gradrho_RtSnrF_H1T3}, similarly to figure~\ref{fig:jpdfH1T3}. In both flows H1 and T3, two regions have high probability density (see arrows I and II). Region I corresponds to the rim where $\cos^2{\zeta^n} + \cos^2{\zeta^r} \approx 1$, i.e.  nearly within the $\hat{\bm n}$--$\hat{\bm r}$ plane and thus  $\bnabla\rho\perp\hat{\bm f}$. Note that $\hat{\bm f} = \hat{\bm n} \times \hat{\bm r}$ and corresponds to the direction of non-rotational straining (shearless) motions. Region II corresponds to the centre where $\cos{\zeta^n} = \cos{\zeta^r} \approx 0$, i.e. $\bnabla \rho \parallel \hat{\bm f}$. These two alignment properties between the $\bnabla \rho$ and $\hat{\bm f}$ indicate two distinct states of mixing, which we will elaborate on later in the paper.

We can gain further insight by conditioning the statistics of $\zeta^n$ and $\zeta^r$ on the shear $S$, as in the previous section. The blue curves in figure~\ref{fig:Gradrho_RtSnrF_H1T3}\textit{a,b} shows that $\cos\zeta^n$ and $\cos\zeta^r$ both become more confined to 0 with high shear (darker blue curves), thus belonging increasingly to region II. This is also evidenced by the joint PDFs conditioned at $k\in(2,2.5]$ (top right insets), showing a strong peak at the origin.  By contrast, the joint PDFs conditioned at $k\in(0,0.5]$ (top left insets) show that $\zeta^n$ and $\zeta^r$ belong to region I. This suggests that $\bnabla \rho\parallel \hat{\bm f}$ in regions of high shear %e.g. isopycnals resemble $\rho=\rho_3$ in figure \ref{fig:RSdecomp}\textit b) 
and $\bnabla \rho\perp \hat{\bm f}$ in regions of low shear. % (e.g. isopycnals resemble $\rho=\rho_1$ or $\rho_2$  in figure \ref{fig:RSdecomp}\textit b). 
 The asymmetry seen in the inserts for the turbulent case (see dashed lines in I and II in figure~\ref{fig:Gradrho_RtSnrF_H1T3}\textit{b}) also suggest that $\bnabla \rho$ is slightly more aligned with $\hat{\bm n}$ than with $\hat{\bm r}$. % (i.e. isopycnals close to $\rho_2$ are more likely). 

Although \cite{Jiang2022} discovered that $\bnabla\rho$ vectors frequently exhibit perpendicular orientation relative to $\hat{\bm r}$ within regions of pronounced stratification, % and the most dominant vortices show a reduced tendency for perpendicularity with $\bnabla\rho$,} 
the present findings highlight the importance of the alignment between $\bnabla \rho$ and the $\hat{\bm n} - \hat{\bm r}$ plane %(and its normal $\hat{\bm f}$) 
and the dependence on shear strength. We recall that in SID flows, %the shear vector $\bm S$ aligns closely with $\bm \omega$, which belongs to the $\hat{\bm n} - \hat{\bm r}$ plane. 
$\hat{\bm n} - \hat{\bm r}$ plane  is (statistically) preferentially inclined by $0-15^\circ$ to the `true horizontal' plane \citep{Jiang2022} , i.e. $\hat{\bm{f}} \parallel \hat{\bm g}$ (gravity). %=-\cos\theta \hat{\bm z}+\sin\theta \hat{\bm x}$ 
In a stably stratified shear layer, we expect preferentially $\bnabla \rho \parallel \hat{\bm g}$ (hydrostatic equilibrium), and therefore $\bnabla \rho \parallel \hat{\bm f}$, %i.e. isopycnals $\rho_3$ in figure \ref{fig:RSdecomp}\textit b, 
which is what is found at high-shear regions where the stratification exerts sufficient strength to mitigate the stirring caused by neighboring vortical structures
However, in regions with weak shear, such as the middle layer of turbulent flow, $\bnabla \rho$ reorients from $\bnabla \rho \parallel \hat{\bm f}$ to $\bnabla \rho \parallel \hat{\bm n} - \hat{\bm r}$ plane (see figure \ref{fig:RSdecomp}), which arises as a result of nearby flow motions that distort local fluid parcels and disrupt isopycnals. In these unstable regions, figure~\ref{fig:Gradrho_RtSnrF_H1T3}(\textit b) suggests that rotation is more effective than non-rotational shear in distorting the isopycnals. We believe that the intricate interplay between these flow motions and buoyancy may affect the local instability due to their differing timescales.% (e.g. $\rho_2$ in figure \ref{fig:RSdecomp}\textit b).  

\subsection{Squared-density-gradient ratios}
\label{sec:Decomp_densitygradient_dissip}

\begin{figure}
\centering
\includegraphics[width=1.0\textwidth]{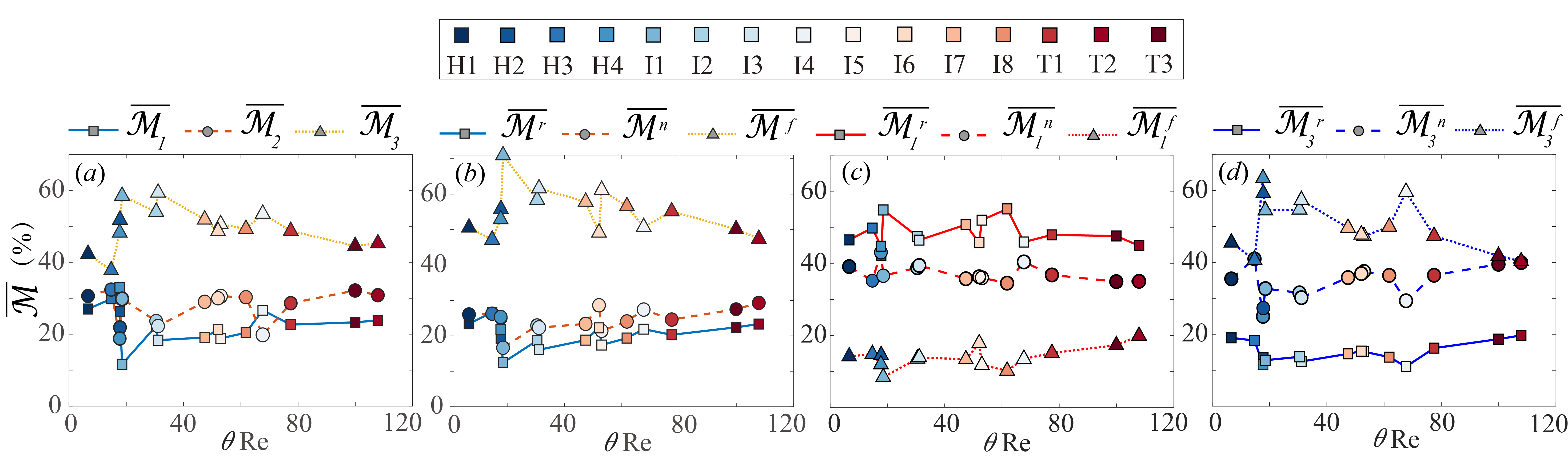}%
\caption{Averaged SDGRs defined in \protect{\eqref{eqn:densityprojectionratioij}} for the 15 experimental datasets. (\textit{a-d}) Variation with $\theta \Rey$ of $\overline{\mathscr{M}_{i}}$ (\textit{a}),  $\overline{\mathscr{M}^\phi}$ (\textit{b}), $\overline{\mathscr{M}_{1}^\phi}$ (\textit{c}),  $\overline{\mathscr{M}}_{3}^\phi$ (\textit{d}). }
\label{fig:dissGradrhoProj1}
\end{figure}

\begin{figure}
\centering
\includegraphics[width=0.86\textwidth]{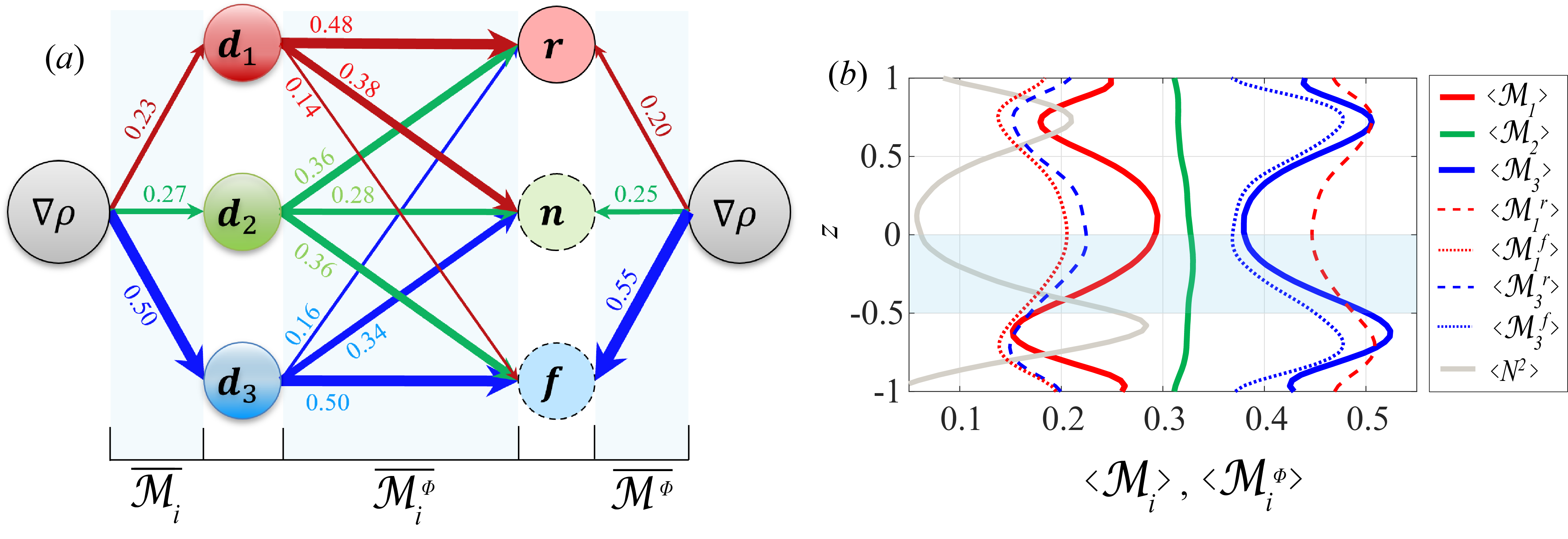}%
\caption{(\textit a) Schematic diagram of overall values of the SDGRs represented by the thickness of arrows and text. (\textit b) Vertical profiles of the seven key SDGRs for the turbulent flow T3, together with the buoyancy frequency $N^2$.  The blue shaded region in (\text d) is re-used in the analysis of figure~\protect{\ref{fig:T3_mix_SDGR}}.}
\label{fig:dissGradrhoProj2} 
\end{figure}

We now examine in figure \ref{fig:dissGradrhoProj1} and figure \ref{fig:dissGradrhoProj2} the alignment of $\bnabla\rho$ in both kinematic bases using the SDGRs defined in \eqref{eqn:densityprojectionratioij}. The overbar $\overline{\mathscr{M}}$ averages an SDGR over the entire available shear-layer volume ($x,y,z$) and time ($t$), as in figure \ref{fig:dissGradrhoProj1}, whereas brackets $\langle\mathscr{M}\rangle(z)$ average in $x$, $y$, and $t$, retaining the $z$ dependence, as in figure \ref{fig:dissGradrhoProj2}(\textit b). 

Figure \ref{fig:dissGradrhoProj1} compares the averaged SDGRs in the 15 flows ranging from H1 to T3, with increasing turbulent intensity controlled  by the product $\theta\Rey$ (where $\theta$ is in radians). In panel (\textit{a}), we find that typically $\overline{\mathscr{M}_3}>\overline{\mathscr{M}_2}>\overline{\mathscr{M}_1}$, confirming the results of \S~\ref{sec:egradrho} that $\bnabla\rho\parallel \hat{\bm d_3}$ is preferential, although this trend of dominant $\overline{\mathscr{M}_3}$ decreases slightly as $\theta\Rey$ increases. For $\overline{\mathscr{M}^\phi}$ in panel (\textit{b}), the distribution is similar to $\overline{\mathscr{M}_i}$ in (\textit{a}). Both $\overline{\mathscr{M}^n}$ and $\overline{\mathscr{M}^r}$ show a nearly linear increase as $\overline{\mathscr{M}^f}$ decreases, indicating that the impact of shear and rortices on density becomes more pronounced with intense turbulence. The reason behind the deviation of $\bnabla\rho$ from $\hat{\bm d_3}$ or $\hat{\bm f}$ %(i.e. the decrease in $\overline{\mathscr{M}_3}$ or $\overline{\mathscr{M}^f}$) 
at higher $\theta\Rey$ can be attributed to more intense mixing and overturning.

%Panels (\textit{a-b}) compare the averaged SDGRs in the 15 flows ranging from H1 to T3, with increasing turbulent intensity controlled  by the product $\theta\Rey$ (where $\theta$ is in radians). 
% The product signals the regime transition and scales with kinetic energy dissipation within the duct \citep{Lefauve2019}. %\AL{Add some more details about meaning of this product.}
%In panel (a), we find that typically $\overline{\mathscr{M}}_3>\overline{\mathscr{M}}_2>\overline{\mathscr{M}}_1$, confirming the results of \S~\ref{sec:egradrho} that $\bnabla\rho\parallel \hat{\bm d_3}$ is preferential, although this trend of dominant $\overline{\mathscr{M}}_3$ decreases slightly as turbulence intensity increases. 

In panel (\textit{c}), we find that $\overline{\mathscr{M}_1^r}>\overline{\mathscr{M}_1^n}>\overline{\mathscr{M}_1^f}$, indicating that the component of the density gradient along the direction of maximum dissipation $\bnabla \rho_{1}$ aligns preferentially with rotation $\hat{\bm r}$, and secondarily with non-rotational shear $\hat{\bm n}$.
As $\overline{\mathscr{M}_1^f} \le 20\%$, it is likely that $d_1$ lies close to the $\hat{\bm n} - \hat{\bm r}$ plane. If $\hat{\bm r}\parallel\hat{\bm d_1}$, both a rotation and stretching occur along the $\hat{\bm d_1}$ direction, strengthening the velocity induced by the non-rotational shear, as depicted in figure \ref{fig:RSdecomp}(\textit a).  
This suggests that the stretching of the $\hat{\bm n} - \hat{\bm r}$ plane, especially along the $\hat{\bm r}$ direction, is closely related to the dissipation.  

%
%%%%%%%%%%%%%%%

% \begin{figure}
% \centering
% \includegraphics[width=1.01\textwidth]{figure/dissipation_tensor_surf3.png}%
% \caption{\label{fig:dissTensorSurf}\revise{Schematic of deformation of dissipation tensor surface due to shearing and rotational structures. (\textit a) Initial dissipation tensor surface (truncated triaxial ellipsoid surface depicted in green) with the length of semi-axis parallel to $\hat{\bm d_i}$ being proportional to $\lambda_i$; the purple ellipsoid is aligned with $\hat{\bm d_1}-\hat{\bm d_2}$ plane;  (\textit b) surface undergoing stretching and rotation along the direction of $\hat{\bm d_1}$, when $\hat{\bm r}\parallel\hat{\bm d_1}$ and $\hat{\bm n}\parallel\hat{\bm d_2}$; (\textit c) surface undergoing stretching and rotation along the direction of $\hat{\bm d_2}$, when $\hat{\bm n}\parallel\hat{\bm d_1}$ and $\hat{\bm r}\parallel\hat{\bm d_2}$.}}
% \end{figure}

In panel (\textit{d}) we see that the preferential alignment is reversed for $\bnabla \rho_3$: $\overline{\mathscr{M}_{3}^f}>\overline{\mathscr{M}_{3}^n}>\overline{\mathscr{M}_{3}^r}$, i.e. primarily aligned with $\hat{\bm f}$. In the most turbulent flows T2 and T3,  $\overline{\mathscr{M}_{3}^f} \approx \overline{\mathscr{M}_{3}^n} \approx 2 \overline{\mathscr{M}_{3}^r}$, i.e. $\bm\nabla \rho_3$ is preferentially in the $\hat{\bm n} - \hat{\bm f}$ plane and perpendicular to $\hat{\bm r}$.  %$\bnabla \rho_3\perp\hat{\bm r}$. 
%The associated isopycnals ($\rho_2$ and $\rho_3$ in figure \ref{fig:RSdecomp}\textit b) are most prone to turbulent mixing. 
We recall from  \S\ref{sec:egradrho}-\ref{sec:align_structure_gradrho} that more stable, high shear regions have approximately $\bnabla \rho \parallel \hat{\bm f} , \hat{\bm d_3} , \hat{\bm g}$, thus $\bnabla \rho_{3}$  is the component of the gradient most likely to be disturbed during mixing. Having $\bnabla\rho_3 \perp \hat{\bm r}$ signals stirring and overturning, a requirement for diapycnal mixing by small-scale diffusion. 

As $\theta$ increases at nearly constant $\Rey$, the $\hat{\bm{d}_3}-\bm{\nabla}\rho$ alignment decreases, as well as the $\hat{\bm{f}}-\bm{\nabla}\rho$ and $\hat{\bm{d}_3}-\hat{\bm{f}}$ alignments.
This is observed in I3, I7, I8, and T3 with $\theta=2^\circ - 6^\circ$ and $\Rey\approx1000$. Conversely, increasing $\Rey$ at constant $\theta=5^\circ$ has a more limited impact on  alignment,  as seen in cases like H2, H4, I5, I8, and T3. %, except for regions influenced by strong vortices as discussed in \citealp{Jiang2022}. 
%Only a modest increased alignment between $\hat{\bm r}$ and $\hat{\bm d_1}$ and a decreased alignment between $\hat{\bm f}$ and $\hat{\bm d_3}$ is observed in H2, H4, I5, I8, and T3. 
This suggests that $\theta$ exerts a more significant influence than $\Rey$ on $\bnabla \rho$ alignment, echoing the findings of \cite{Lefauve20221} that $\theta$ increases overturning faster than $Re$.

%Our finding that the rotation axis is found to be the least aligned with this density gradient ($\overline{\mathscr{M}}_{3r}$ being minimum). 
%It is also easy to see that when $\hat{\bm r}$ aligns with the density gradient, rotation alone cannot generate any diapycnal mixing, but it can enhance diffusion within the density isosurface. 
%For the shearing deformation, figure \ref{fig:dissGradrhoProj}(\textit c) reveals  that he $\overline{\mathscr{M}}_{3n}$ is approximately twice the value of $\overline{\mathscr{M}}_{3r}$, and both tend to increase at higher $\theta\Rey$.  These results suggest that, \result{$\hat{\bm d_3}$ aligns more frequently with $\hat{\bm n}$ than with $\hat{\bm r}$, and in turbulent cases, the likelihood of $\hat{\bm d_3}$ aligning with the $\hat{\bm n} - \hat{\bm r}$ plane increases.} 

Figure \ref{fig:dissGradrhoProj2}(\textit a) summarises the typical averaged SDGRs across all 15 flows and encapsulates the typical geometry of stratified turbulence in our experimental data. The left-hand branches (i.e. the three arrows under the blue shading indicated by $\overline{\mathscr{M}_i}$) show the three `dissipation' SDGRs, the right-hand branches show the three `structural' SDGRs (indicated by $\overline{\mathscr{M}^\phi}$), and the middle branches show the nine `mixed' SDGRs indicated by $\overline{\mathscr{M}_i^\phi}$. Note the approximate symmetry between the left- and right-hand branches, indicating a robust relation between diapycnal transport, structural and dissipation coordinates.

Finally, figure \ref{fig:dissGradrhoProj2}(\textit{b}) shows the variation of seven key SDGRs along $z$ (in the duct's frame of reference) in the turbulent flow T3. To examine the effect of stratification, we superimpose the averaged nondimensional buoyancy frequency $\langle N^2\rangle=-Ri_b\partial_z \rho$  (thick grey line). 
We find that $\langle{\mathscr{M}}_{1}\rangle$ (thick red line) is maximum in the middle layer where $\langle N^2\rangle$ is minimum, indicating that $\bnabla \rho$ is more distorted and thus more aligned with $\hat{\bm d_1}$ in the turbulent mixing layer. Vice versa, $\langle \mathscr{M}_{1}\rangle$ is minimum at the upper and lower density interfaces of this mixing layer where $\langle N^2\rangle\gtrsim 0.2$. This trend is exactly reversed for $\langle{\mathscr{M}}_{3}\rangle$, which reaches a minimum value in the well-mixed region and a maximum value at the interfaces.  Furthermore, $\langle\mathscr{M}_{2}\rangle$ is relatively uniform in $z$ and independent of the stratification. We also observe the same pattern of reversed behaviour between $\langle\mathscr{M}_1^r\rangle$ and $\langle\mathscr{M}_1^f\rangle$, as well as between $\langle\mathscr{M}_3^r\rangle$ and $\langle\mathscr{M}_3^f\rangle$. Recalling from   \S\ref{sec:align_structure_gradrho} that $\bnabla\rho\perp \hat{\bm r}$ at high shear, the finding here that stronger stratification favours the alignment of $\hat{\bm r}$ and $\hat{\bm d_1}$ (larger $\langle\mathscr{M}_1^r\rangle$), which is consistent with the fact that more diapycnal transport occurs at the interface. This agrees with \cite{Jiang2022} who studied the interaction of hairpin-like structures and density gradients through $\bm R\times\bnabla\rho$, as well as with \cite{Riley2023}, who found that most potential energy dissipation occurs near density interfaces.  %They discovered that the most significant interactions between the hairpin-like structure and the density gradient take place in close proximity to the upper and lower interfaces of the stratified layer.
%The plot illustrates that \result{stratification significantly affects the alignment  }%between the dissipation direction and the structural coordinate}.   %\AL{What do you mean by this? It's vague. } \AL{What physics do we extract from this section?}

%%%%%%%%%%%%%%%% 
\subsection{Link with mixing coefficient} 
\label{sec:ImplicationMix}

\begin{figure}
\centering
\includegraphics[width=0.98\textwidth]{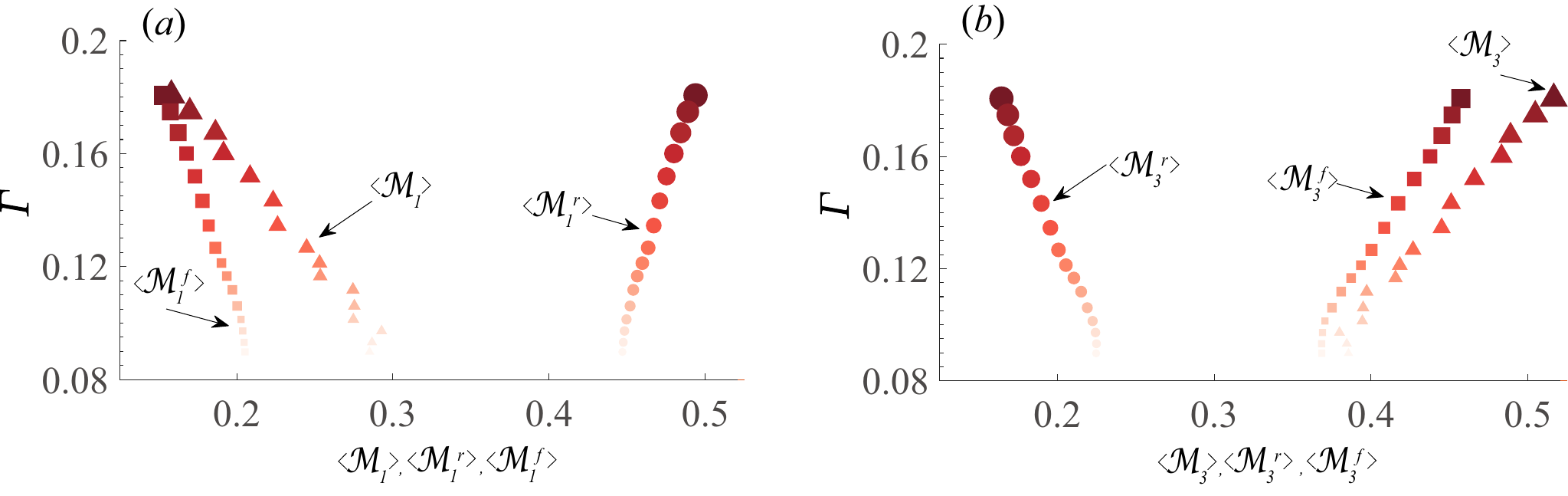}
\caption{Correlation between the turbulent flux coefficient $\Gamma$ and the six SDGRs (a) $\langle\mathscr{M}_{1}\rangle, \langle\mathscr{M}_{1}^r\rangle, \langle\mathscr{M}_{1}^f\rangle$, (b)  $\langle\mathscr{M}_{3}\rangle, \langle\mathscr{M}_{3}^r\rangle, \langle\mathscr{M}_{3}^f\rangle$  in the T3 case within the shear layer ($z\in[-0.5,0]$). The size of symbols indicates $|z|$, and the darker colours indicate higher stratification $N^2$.} 
\label{fig:T3_mix_SDGR}
\end{figure}

Figure \ref{fig:T3_mix_SDGR} investigates the relationship between the SDGRs and the flux coefficient $\Gamma\equiv \bar{\mathcal{B}}/\bar{\epsilon'}$, the ratio of the globally-averaged turbulent buoyancy flux $\mathcal{B} \equiv Ri_b\,  w'\rho'$ and turbulent kinetic energy dissipation $\epsilon' \equiv (2/Re)\, e'_{ij} e'_{ij}$ where the perturbations (prime variables) are taken with respect to the $(x,t)$ average \citep[\S~2.2]{Lefauve20222}. %\AL{Why did you choose those 6 SDGRs and not others?} % \XJ{because we observe those in fig4(e) and M2 and M_i_n is nearly constant along z}
We restrict the analysis to the T3 case within the active turbulent lower layer $z\in[-0.5,0]$ (blue shaded region in figure \ref{fig:dissGradrhoProj2}\textit e). %Additionally, the plot includes a comparison of the flux coefficient $\Gamma_\chi$, which is defined as the ratio of scalar variance to dissipation $\langle\mathcal{\chi}\rangle/\langle\varepsilon\rangle$ (represented by open symbols on the right y-axis).
Larger symbols indicate higher $|z|$, and darker colors indicate high $N^2$ (in this region, $N^2\propto |z|$).

First, the linear relations between $\Gamma$ and the SDGRs suggest a clear link between the efficiency of mixing, approximated by this flux coefficient \citep{CaulfieldPRF2020}, and the geometry encapsulated in the SDGRs. Second, $\Gamma$ decreases in proportion to %inversely proportional to 
$\langle\mathscr{M}_1\rangle=\cos^2\angle (\bnabla\rho,\hat{\bm d_1})$ and $\langle\mathscr{M}_1^f\rangle$ but increases in proportion to $\langle\mathscr{M}_{1}^r\rangle$ (panel (\textit{a})). 
Looking at the direction of minimal dissipation, the opposite is observed: $\Gamma$ is proportional to $\langle\mathscr{M}_3\rangle=\cos^2\angle (\bnabla\rho,\hat{\bm d_3})$ and $\langle\mathscr{M}_3^f\rangle$ but inversely proportional to $\langle\mathscr{M}_{3}^r\rangle$. We interpret this by the expectation that low $\mathscr{M}_1$ or high $\mathscr{M}_3$ indicates the alignment of isopycnals along the direction of greatest dissipation (or stretch, see \S~\ref{sec:DensityDecomp}), and hence an increase in the surface area of isopycnals and diffusive mixing.
Third, figure \ref{fig:T3_mix_SDGR} shows that closer to the density interface (larger darker symbols), $\hat{\bm r}$ aligns more with $\hat{\bm d_1}$ (i.e. $\langle\mathscr{M}_{1}^r\rangle$ becomes larger), while  $\bnabla \rho$ aligns more with $\hat{\bm f}$ and $\hat{\bm d_3}$ (i.e. $\langle\mathscr{M}_{3}^f\rangle$ and $\langle\mathscr{M}_{3}\rangle$ become larger). This results in the $\hat{\bm n} - \hat{\bm r}$ plane aligning more with the $\hat{\bm d_1}-\hat{\bm d_2}$ plane, indicating stronger overturning and stretching leading to mixing.

%%%%%%%%%%%

\section{\label{sec:Con}Conclusions} 

In summary, we have explored the alignment of the density gradient $\bnabla \rho(\bm x,t)$ in two local, instantaneous orthogonal bases determined from flow kinematics. Our goal was to shed light on the geometry of shear-driven, stably-stratified turbulent mixing in a stratified inclined duct experiment. We used the datasets of increasingly turbulent flows investigated in \cite{Jiang2022}, with a focus on two main flows representative of the Holmboe wave and turbulent regimes. %Our main results are summarised by figure \ref{fig:RSdecomp}, figure \ref{fig:dissGradrhoProj}(d-e), and figure \ref{fig:T3_mix_SDGR}

The first kinematic basis $(\hat{\bm n}, \hat{\bm r}, \hat{\bm f})$, constructed using the non-rotational shear $\hat{\bm n}$ and rigid-body rotation $\hat{\bm r}$, showed that in regions of weak shear, $\bnabla\rho$ preferentially aligns in the $\hat{\bm n} - \hat{\bm r}$ plane. % (in regions of weak shear) or normal to it (in high shear). 
However, increasing shear strength tends to reorient $\bnabla \rho$ normal to $\hat{\bm n} - \hat{\bm r}$ plane, thus increasing the potential for destruction of stratification by rotational and shearing structures. The second kinematic basis  $(\hat{\bm d_1}, \hat{\bm d_2}, \hat{\bm d_3})$, constructed using the principal directions of the pseudo-dissipation tensor \eqref{eq:pseudo-diss-tensor} 
showed that $\bnabla\rho$ aligns preferentially with the direction of minimum dissipation $\hat{\bm d_3}$ and thus normal to the maximum dissipation $\hat{\bm d_1}$, a trend that is exacerbated in regions of high shear.

The squared-density-gradient ratios (SDGRs, see \eqref{eqn:densityprojection1i}-\eqref{eqn:densityprojectionratioij}) quantified the alignment of $\bnabla \rho$ in both kinematic bases to reveal that about $\overline{\mathscr{M}_{1}}\approx 23\%$ of the total squared-density-gradient ratio is along the maximum dissipation $\hat{\bm d_1}$, which is itself dominated by flow structures in the $\hat{\bm n} - \hat{\bm r}$ plane. By contrast, a ratio $\overline{\mathscr{M}}_{3}\approx 50\%$ is along the minimum dissipation $\hat{\bm d_3}$, which aligns preferentially perpendicular to $\hat{\bm n}- \hat{\bm r}$ and makes it more susceptible to being distorted through overturning by vortices and hence more diapycnal mixing. 

Focusing on the most turbulent flow, the variation of SDGRs across the shear layer explained why more efficient mixing, quantified by $\Gamma$, occurs at the edges of the mixing layer. At the edges (stronger stratification), $\hat{\bm r}$ aligns more with the maximum dissipation $\hat{\bm d_1}$, and $\bnabla \rho $ aligns more with the minimum dissipation $\hat{\bm d_3}$, and therefore, prone to overturning by rotational structures. The $\hat{\bm n} - \hat{\bm r}$ plane becomes more aligned with the $\hat{\bm d_1}-\hat{\bm d_2}$ plane, enhancing stirring and overturning and increasing mixing efficiency. This mechanism explains the robust linear correlations between the mixing efficiency $\Gamma$ and the SDGRs shown in figure \ref{fig:T3_mix_SDGR}. Our findings also rationalise that $\Gamma$ decreases with increasing %is inversely proportional to 
$\mathscr{M}_{1}$ and increases with %proportional to 
$\mathscr{M}_{3}$ because isopycnals aligned along the direction of maximum dissipation %(or stretch)
are more exposed to stretching and diffusion. These insights into the geometry of the density gradient and dissipation constitute a step towards a better physical understanding of turbulent mixing in stratified shear flows. Future research should study the applicability of these findings to stratified flows that differ from the present stably-stratified, shear-driven inclined duct flows.

\vspace{0.3cm}

The authors acknowledge the ERC Grant No 742480 `Stratified Turbulence And Mixing Processes'. A.L. acknowledges a Leverhulme Early Career Fellowship and a NERC Independent Research Fellowship (NE/W008971/1). 
\vspace{0.3cm}

%{\textbf{Declaration of interests}} \par
Declaration of Interests. The authors report no conflict of interest.

\bibliographystyle{jfm}
\bibliography{localized_GDP}

%
% \begin{figure}
% \centering
% \includegraphics[width=0.5\textwidth]{figure/Abstract1.png}\caption*{\label{fig:Abstract}Graphic abstract at the ratio of 1.2:1.}
% \end{figure}
%

\end{document}